\documentclass[fleqn,10pt]{olplainarticle}
\usepackage{subcaption}
\title{Differentiation of Wild-Type and CRISPR-Modified Colon Cancer Cells Using Brillouin Microscopy}

\author[1]{Mykyta Kizilov}
\author[1]{Vsevolod Cheburkanov}
\author[1]{Vladislav Yakovlev}
\affil[1]{Texas AM University, Department of Biomedical Engineering, College Station, TX 77843, USA}

\begin{abstract}
This study investigates the mechanical properties of colon cancer cells through Brillouin microscopy, focusing on the differentiation between wild-type (WT) and CRISPR-modified cells. Brillouin microspectroscopy, a non-invasive technique, was employed to measure Brillouin shifts and full width at half maximum (FWHM) values of the cells in vitro. Using a custom-built confocal Brillouin microspectrometer, both WT and CRISPR-modified cells exhibited distinct mechanical responses. Statistical analysis revealed that WT cells had different stiffness and viscosity compared to CRISPR-modified cells, as indicated by their Brillouin shift and FWHM values. The data suggest that Brillouin spectroscopy offers a viable method to differentiate between normal and mutated cells at the subcellular level, providing new insights into cellular mechanical properties relevant to cancer research. These findings hold potential for advancing non-invasive diagnostic techniques and understanding cellular mechanics in oncology.

\end{abstract}

\begin{document}

\flushbottom
\maketitle
\thispagestyle{empty}

\section{Introduction}
Colon cancer, also known as colorectal cancer (CRC), remains a significant public health challenge in the United States and in the world \cite{araghi2019}. It is the second most common cause of cancer-related deaths, with approximately 153,020 individuals expected to be diagnosed and 52,550 deaths anticipated in 2023 alone. The burden is particularly concerning for younger adults \cite{oconnell2003}, as the incidence among those under 50 has been rising steadily by 1–2\% annually since the mid-1990s, reversing the declining trend seen in older populations \cite{siegel2023}. These trends underscore the urgent need for improved detection and intervention strategies.

Despite advances in screening and early detection, disparities in diagnosis and treatment persist across socioeconomic and demographic groups. For instance, African Americans and Alaska Natives experience disproportionately higher incidence rates and mortality from CRC compared to other racial and ethnic groups \cite{liss2014,singh2017}. Additionally, rural and economically disadvantaged populations often face limited access to high-quality screening programs, exacerbating health inequities \cite{miller2019,egeberg2008}. Addressing these disparities is crucial for reducing the overall burden of the disease.

Efforts to combat colon cancer have included widespread screening campaigns. Colonoscopy and sigmoidoscopy are recognized as gold-standard procedures that reduce incidence and mortality by enabling the removal of precancerous lesions. However, only 56\% of at-risk adults aged 50 and older reported undergoing a colonoscopy or sigmoidoscopy as of 2006, leaving a substantial portion of the population vulnerable \cite{stock2010}. This underutilization of screening emphasizes the need for enhanced public health strategies.

Recent research underscores the growing economic burden associated with CRC, especially as the U.S. population ages. Hospitalization charges for colon cancer exceeded \$4.57 billion annually in the 1990s \cite{mingsi2022} and are projected to increase dramatically as the elderly population expands. This situation highlights the urgency for cost-effective preventive strategies and early detection programs \cite{seifeldin1999}.

As a result, colon cancer presents a pressing healthcare challenge in the U.S., marked by rising incidence in younger populations, persistent racial and socioeconomic disparities, and significant economic impact. Increased public awareness, equitable access to screening, and innovative diagnostic technologies are critical to addressing this multifaceted burden.

Building upon these public health challenges, recent advancements in genetic research have enabled the study of both wild-type (WT) and CRISPR-modified colon cancer cells. These models offer valuable insights into tumor biology, therapeutic vulnerabilities, and novel treatment strategies, thereby bridging clinical challenges with molecular understanding.

The importance of wild type colon cancer models lies in their ability to provide a more accurate representation of the human disease, facilitating the study of cancer progression and the development of therapeutic strategies. For example, models such as decellularized mouse colons offer a physiologically relevant extracellular matrix that supports the growth and differentiation of human cancer cells, allowing researchers to study the influence of the extracellular matrix on cancer progression \cite{Alabi2019Decellularized}. Additionally, wild type models are crucial for understanding the genetic and molecular mechanisms underlying colon cancer; they enable the identification of cancer driver genes and the study of gene interactions that contribute to tumorigenesis3. Furthermore, these models are instrumental in the development of new therapeutic approaches, as they allow for the testing of potential treatments in a controlled environment that closely mimics human colon cancer \cite{Chen2016A, Roper2017In}. Overall, wild type colon cancer models are indispensable tools in advancing our understanding of colon cancer biology and improving patient outcomes.

In parallel, the role of wild-type colon cancer in therapy development is pivotal, particularly in the context of genetic mutations and their impact on treatment efficacy. Wild-type KRAS status in colon cancer is crucial for the effectiveness of therapies targeting the epidermal growth factor receptor (EGFR). Patients with KRAS wild-type tumors are more likely to respond to anti-EGFR monoclonal antibodies, making KRAS mutation testing a standard procedure before initiating such treatments \cite{Parsons2013Personalized}. Additionally, the presence of wild-type p53 in colon cancer cells has been shown to enhance the cytotoxic effects of therapeutic agents like irradiation and 5-fluorouracil, suggesting that the loss of wild-type p53 may contribute to resistance against these treatments \cite{Yang1996Wild-type}. Furthermore, the introduction of wild-type p53 through gene therapy has demonstrated potential in suppressing tumor growth and angiogenesis, offering a promising strategy for colon cancer treatment \cite{Ogawa1997Novel, Bouvet1998Adenovirus-mediated}. The wild-type isocitrate dehydrogenase 2 (IDH2) also plays a protective role against oxidative DNA damage in colorectal cancer, highlighting its potential as a therapeutic target6. Collectively, these findings underscore the importance of understanding the genetic landscape of colon cancer, particularly the roles of wild-type genes, in developing effective therapeutic strategies.

In addition to wild-type models, colon cancer is characterized by a complex array of genetic mutations that drive its progression. The advent of CRISPR-Cas9 gene editing technology has revolutionized the study and treatment of colon cancer by enabling precise modifications of the genome. This technology allows researchers to identify and manipulate oncogenes, tumor suppressor genes, and drug-resistance genes, thereby facilitating the development of personalized therapeutic strategies \cite{Meng2023Application, Sokhangouy2024Recent}. CRISPR-Cas9 has been instrumental in creating experimental models and conducting genome-wide screenings, which are crucial for understanding the genetic underpinnings of colon cancer and for developing targeted treatments \cite{Sokhangouy2024Recent, Michels2020Pooled}. Furthermore, innovative delivery systems, such as plant-derived exosome-like nanoparticles, are being explored to enhance the safety and efficiency of CRISPR-based therapeutics, specifically targeting cancer cells in the colon \cite{Hillman2023The}. The integration of CRISPR technology with organoid models has also provided insights into the origins of mutational signatures in cancer, offering potential diagnostic and prognostic applications \cite{Drost2017Use, Ramakrishna2021Application}. Overall, CRISPR-Cas9 continues to be a transformative tool in the fight against colon cancer, paving the way for advancements in genetic research and personalized medicine.

These advancements have further propelled CRISPR technology to the forefront of colon cancer research, offering new therapeutic insights and future research directions. CRISPR/Cas9 systems have been instrumental in identifying genetic vulnerabilities and therapeutic targets in colon cancer models. For instance, genome-wide CRISPR knockout screens in colon cancer organoids have revealed specific gene dependencies that can be exploited for targeted therapies, such as TP53-dependent vulnerabilities and pathways like the WNT and BRAF pathways \cite{Buckhaults2023Abstract}. Additionally, CRISPR technology has facilitated the development of experimental models that mimic the genetic landscape of colon cancer, enabling the discovery of novel oncogenes and drug-resistance genes \cite{Sokhangouy2024Recent, Meng2023Application}. These models are crucial for understanding the complex genetic variations in colon cancer and for developing personalized treatment strategies. Furthermore, CRISPR screens have identified cholesterol biosynthesis as a potential therapeutic target, highlighting the role of metabolic pathways in cancer stemness and drug resistance \cite{Gao2021CRISPR}. As research progresses, the focus will be on validating these targets and integrating CRISPR-based approaches into clinical practice to enhance the precision and efficacy of colon cancer treatments \cite{Sahranavard2023The, Yin2019CRISPR}. Thus, the future of CRISPR in colon cancer research holds promise for more effective and personalized therapeutic interventions.

Complementary to these genetic approaches, optical techniques have emerged as valuable tools in cancer diagnostics. Brillouin spectroscopy is a powerful optical technique used to probe the mechanical properties of materials at a microscopic scale. It is based on the inelastic scattering of light, known as Brillouin scattering, which occurs when photons interact with acoustic phonons in a material. This interaction provides insights into the material's elasticity and viscoelasticity, making it a valuable tool in various fields, including biomedical sciences and material research \cite{Palombo2019Brillouin, Riobóo2021Brillouin}. Recent advancements have enhanced the capabilities of Brillouin spectroscopy, such as the development of multi-wavelength excitation methods to correct background distortions and improve measurement accuracy \cite{Troyanova-Wood2021Multi-Wavelength}. Additionally, innovations like dispersive coherent Brillouin scattering spectroscopy have significantly increased signal acquisition speed, enabling real-time mechanical imaging \cite{Ishijima2021Dispersive}. The technique's non-invasive nature and high spatial resolution have made it particularly promising for applications in clinical diagnostics and the study of biological tissues \cite{Riobóo2021Brillouin, Ballmann2019Nonlinear}. Despite its potential, challenges remain in improving the sensitivity and specificity of Brillouin spectroscopy for certain applications, such as differentiating stages of keratoconus in corneal studies \cite{Seiler2019Brillouin}. Continued research and technological advancements are expected to further expand its applications and effectiveness in various scientific domains.

Furthermore, Brillouin spectroscopy has emerged as a powerful tool in the study of biological systems, offering unique insights into the mechanical properties of biological materials \cite{2025CheburkanovBrillouinSPIE, 2025CheburkanovGliaSPIE}. This technique leverages Brillouin light scattering, which involves the interaction of light with acoustic phonons, to provide a non-contact, label-free method for probing the viscoelastic properties of tissues and cells at a microscopic scale \cite{Palombo2019Brillouin, Zhang2021Mapping, Ballmann2019Nonlinear}. Recent advancements—such as integration with confocal microscopes for high-resolution, three-dimensional mapping of mechanical properties \cite{Zhang2021Mapping} and the use of adaptive optics to improve precision even in complex biological samples \cite{Edrei2018Brillouin}—have further enhanced its applicability. The development of nonlinear and quantum-enhanced Brillouin techniques has expanded its potential, enabling more sensitive and accurate assessments of biomechanical changes in living specimens \cite{Ballmann2019Nonlinear, Li2021Quantum-enhanced}. These improvements underscore the growing importance of Brillouin spectroscopy in biomedical research by providing critical insights into the mechanical underpinnings of biological processes and diseases \cite{Elsayad2019Brillouin}.

In the realm of cancer research, Brillouin spectroscopy offers a non-invasive method to assess the mechanical properties of tissues and cells, which are closely linked to disease physiology and pathology. For instance, studies have demonstrated its capability to differentiate melanoma from healthy tissues based on elasticity, with cancerous tissues generally exhibiting greater stiffness \cite{Troyanova-Wood2019Differentiating, Troyanova-Wood2016Elasticity-based}. Additionally, Brillouin spectroscopy has been successfully applied in imaging cancer metastasis, revealing variations in cell stiffness that correlate with environmental conditions and metastatic potential \cite{Cheburkanov2022Imaging}. The integration of Brillouin spectroscopy with other techniques, such as Raman spectroscopy \cite{2025HarringtonDUVSPIE, 2025HarringtonDUVChemPhysChem, 2025KizilovRamanSPIE}, further enhances its utility by allowing simultaneous biomechanical and biochemical characterization of cancer cells, as shown in glioblastoma research \cite{Rix2022Correlation}. These advancements underscore the potential of Brillouin spectroscopy not only in identifying tumor margins and monitoring tumor progression but also in contributing to a broader understanding of cancer mechanobiology and treatment responses \cite{Nikolić2018Noninvasive}.

Finally, Brillouin spectroscopy offers a promising approach for colon cancer research due to its ability to provide detailed insights into the mechanical properties of biological tissues. As part of a broader coherent spectroscopy approach, this technique allows for the early detection of oncological diseases by analyzing the transmission spectra of cancerous cells, such as human colorectal adenocarcinoma cells (HT-29) \cite{Moguilnaya2018Using}. The integration of stimulated Brillouin scattering with other diagnostic methods enhances the accuracy of identifying tumor markers, achieving a 95\% probability of initial identification. Consequently, this capability is crucial for developing devices aimed at monitoring cancer at its early stages, potentially improving early diagnosis and treatment outcomes for colon cancer patients \cite{Moguilnaya2018Using}.

\section{Experiment design}
Two sets of cell samples were prepared for imaging using a custom-built confocal Brillouin microspectrometer. To mitigate confirmation bias, the specific stiffness values of the cell samples were deliberately withheld. The first sample was labeled "RKo P7 9/10/24 AC P8 9/12/24," indicating its age and condition, while the second sample was labeled "RKo mutant DMEM P7 9/13/24 KM."

The first sample was imaged with two successive objectives: the Nikon Plan Fluor CFI60 20X objective (numerical aperture [NA] 0.5) and the Nikon Fluor CFI60 60X water-immersion objective (NA 1.0). Both objectives were infinity-corrected for 0.17 mm thick cover glass; therefore, variations in flask material thickness had a negligible impact on the effective NA for our applications.

Both samples were imaged within cell culture flasks, where the cells were adhered to the bottom and submerged in nutrient media. The initial imaging run was designed as a test study to evaluate the capability of the system to accurately capture images of cells in this configuration, comparing the quality of widefield and confocal Brillouin channels' focal plane matching.

\subsection{Sample preparation}

\subsection{Setup description and acquisition settings}

\begin{figure*}[h]
    \centering
    \includegraphics[width=1\linewidth]{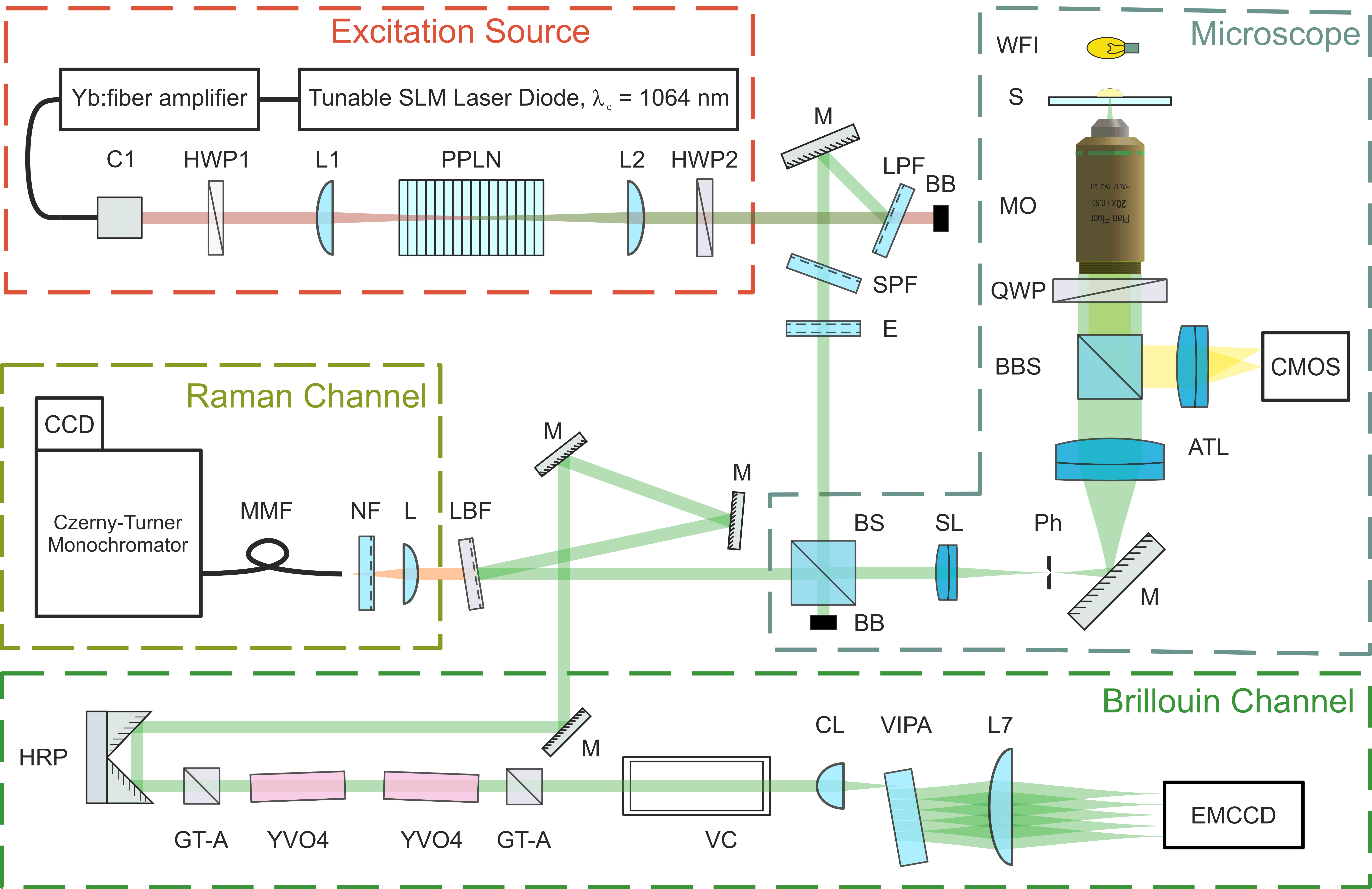}
    \caption{Brillouin confocal microspectrometer layout. \textbf{C} - fiber collimator, \textbf{HWP} – half-wave plate, \textbf{PPLN} - periodically poled lithium niobate second harmonic generation crystal, \textbf{BB} – beam block, \textbf{LPF} - long-pass filter, \textbf{M} - mirror, \textbf{SPF} - short-pass filter, \textbf{BS} – polarizing beamsplitter cube, \textbf{SL} - achromatic doublet scan lens, \textbf{Ph} - precision pinhole, \textbf{ATL} - achromatic tube lens (200mm EFFL), \textbf{BBS} - broadband beamsplitter in quick-insert mount, \textbf{QWP} – quarter-wave plates, \textbf{MO} – microscope objective lens, \textbf{S} – sample, \textbf{WFI} - widefield attachment illumination, \textbf{CMOS} - 5MP camera with a complimentary metal-oxide-semiconductor sensor, \textbf{LBF} - 532nm Bragg grating notch filter, \textbf{NF} - hard-coated interference notch filter, \textbf{MMF} - single core multimode fiber, \textbf{CCD} - camera with a charge-coupled device sensor, \textbf{HRP} - hollow-roof prism, \textbf{GT-A} - Glan-Taylor polarizer, \textbf{YVO4} - non-doped yttrium orthovanadate crystal, \textbf{VC} – Iodine vapor cell, \textbf{CL} – cylindrical lens, \textbf{VIPA} – Virtually Imaged Phase Array, \textbf{EMCCD} - CCD with electron multiplying amplification capabilities}
    \label{fig:br}
\end{figure*}

Elasticity data were acquired using a custom-built upright confocal Brillouin microspectrometer configured in backscattering geometry. The schematic of this setup is depicted in Figure \ref{fig:br}. The system comprises four primary subassemblies: an excitation source, a microscope, a confocal pinhole assembly, and the custom-built Brillouin microspectrometer.

\subsubsection{Excitation}

A custom-built 532 nm laser with a linewidth of less than 1 MHz was utilized as the excitation source. This wavelength was produced as the second harmonic of 1064 nm laser radiation within a periodically poled LiNbO3:MgO crystal (Covesion Ltd.). The 1064 nm light was generated by a tunable single longitudinal mode laser diode (Koheras Adjustik Y10, NKT Photonics) and subsequently amplified using an Yb-doped fiber amplifier (Koheras Boostik HPA Y10, NKT Photonics).

The desired wavelength of 532 nm was effectively isolated from the residual 1064 nm pump light through the use of a combination of long-pass (LPF) and short-pass (SPF) filters. Additionally, a Fabry-Perot etalon, characterized by a free spectral range of 30 GHz and a finesse exceeding 30, was incorporated into the excitation pathway to serve as a sub-GHz laser cleanup filter. This configuration enhanced the purity of the excitation light by minimizing unwanted spectral components, thereby optimizing the overall performance of the system. 

\subsubsection{Microscope}

The microscope body was designed and constructed using standard optomechanical components (Thorlabs). The laser power delivered to the sample was regulated through the use of a 532 nm half-wave plate in conjunction with a polarizing beamsplitter cube (BS, Thorlabs PBS251). A polymer quarter-wave plate (QWP) was employed to achieve orthogonal polarization between the incident and scattered light, transmitting the scattered light through the BS towards the spectrometer for Brillouin and Raman channel detection.

To ensure optimal beam expansion, the laser beam was enlarged to a diameter of 14 mm at $1/e^2$ using a Keplerian telescope beam expander. A precision pinhole was positioned at the intermediate image plane of the telescope, serving to clean up the excitation beam and provide spatial filtration of the collected signal. The microscope was designed to be compatible with the Nikon CFI60 objectives, utilizing a tube lens (ATL) with a focal length of 200 mm. The microscope objective (MO) delivered approximately 5 ± 0.3 mW of 532 nm radiation to the sample while simultaneously collecting the scattered photons.

Target acquisition within the field of view was facilitated by a camera (CMOS) coupled to the microscope with a 200 mm achromatic lens. Transparent specimens were illuminated in transmission mode using a fiber-coupled 6000K LED (Mightex), designated as WFI in Fig. \ref{fig:br}. The position of the sample within the field of view was precisely adjusted via a microscope stage equipped with sub-micron precision. A MCL Nano-LPS stage was used for fine and a custom MCL MicroStage for coarse positioning (Mad City Labs).

\subsubsection{Confocal pinhole selection}

The confocal pinhole was selected to transmit no more than 1 Airy Unit (AU), thereby leveraging the advantages of confocal microscopy, including optical sectioning of the sample and isolation of data from individual layers. To validate pinhole size selection, a broadband dielectric mirror was precisely positioned in the object plane of the MO, and its axial position was systematically scanned. During this process, we measured the throughput of the pinhole and visually assessed beam quality.

Our measurements indicated that at the peak approximately 75\% of the signal was transmitted through the chosen pinhole, and the observed light pattern closely resembled the Fraunhofer diffraction pattern on the power meter. This observation suggests that the selected pinhole diameter was in fact less than 1 AU for the given system configuration, confirming it as a suitable design choice.

\subsubsection{Custom-built Brillouin spectrometer}

The experimental setup involved the detection of both Raman and Brillouin scattered photons, which were effectively separated using a Bragg volumetric grating notch filter optimized for a 532 nm excitation wavelength. This approach facilitated the distinct separation of Brillouin and Raman signals, while also enabling verification of system alignment and beam collimation, with the latter being influenced by the underlying principles of the Bragg grating design and operation.

The filtered signal was coupled into the entrance pupil of a custom-designed Brillouin spectrometer. To mitigate the influence of elastically scattered photons, an iodine vapor cell (Thorlabs) was employed, maintained at a temperature of  $70 ^\circ C$. By implementing a double-pass beam propagation configuration and precisely tuning the laser output wavelength to coincide with the strongest absorption band of molecular iodine, suppression ratios exceeding 40 dB were achieved. The optimal absorption wavelength was determined to be 531.9363 nm, corresponding to line 638, with a wavenumber of 18,799.244 $cm^{-1}$. 

In instances where iodine suppression was inadequate, a narrow-band polarization filter was incorporated into the system. This filter configuration utilized two Glan-Taylor (GT-A) prisms serving as a polarizer and analyzer, respectively, along with a pair of yttrium orthovanadate crystals functioning as long polarization rotators. The dimensions and alignment of the crystals were meticulously adjusted to minimize the transmission of elastically scattered photons by rotating the polarization plane of the elastically scattered signal to a configuration orthogonal to that of the analyzer.

The Brillouin scattering signal was analyzed using a high-dispersion, custom-built single-stage VIPA spectrometer. The VIPA (OP-6721-3371-2, Lightmachinery Inc.) was specifically optimized for 532 nm and featured a free spectral range of 29.98 GHz (1 $cm^{-1}$). Signal spectra were recorded with a water-cooled EMCCD camera (Andor Newton 970P, Oxford Instruments).

\subsection{Signal Pre-Processing}
\begin{figure}[h!]
    \centering
    \includegraphics[width=0.9\linewidth]{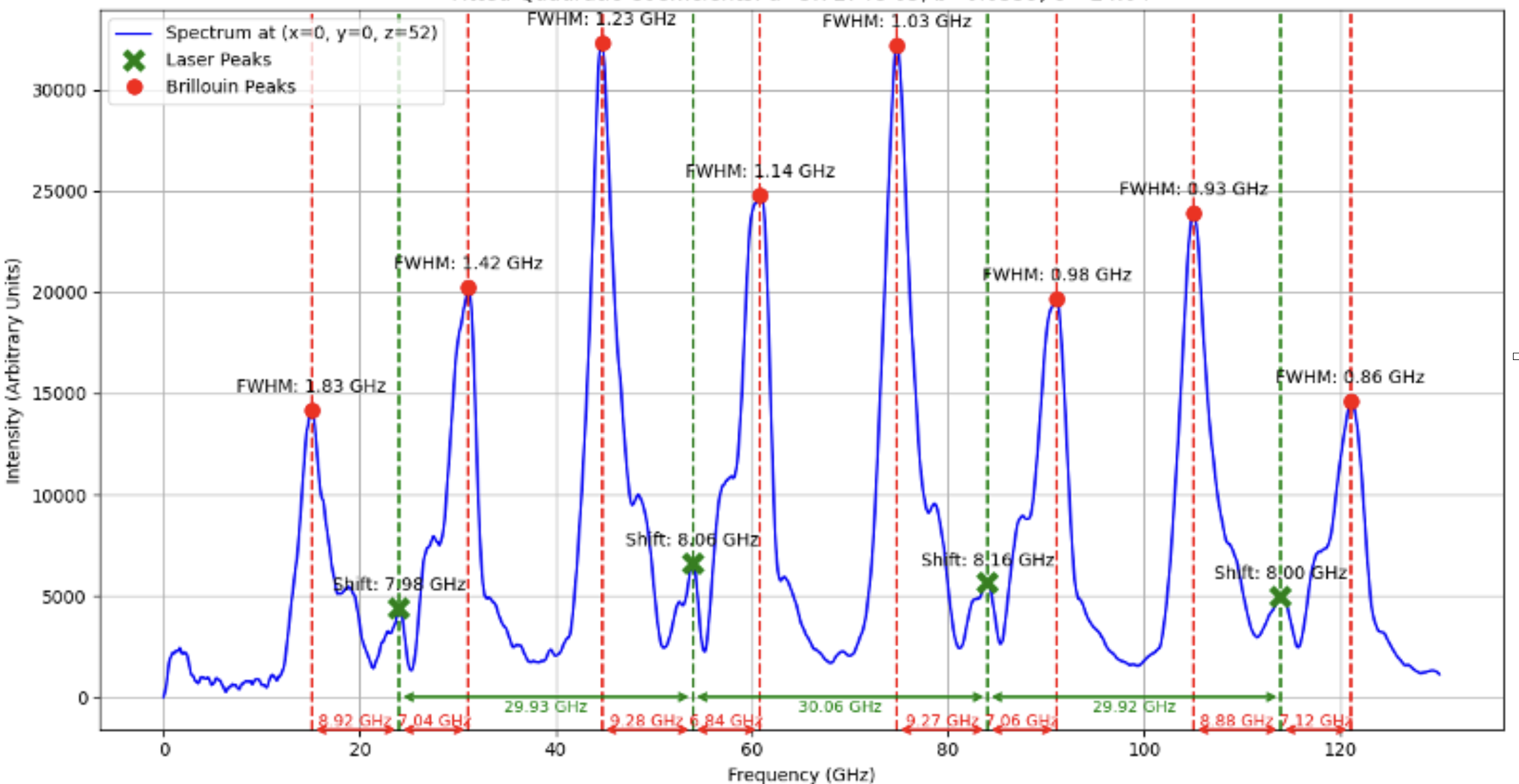}
    \caption{Processed signal for a single pixel}
    \label{fig:Brillouin_spectrum}
\end{figure}

\begin{figure}[h!]
  \centering
  \begin{subfigure}[t]{0.45\textwidth}
    \centering
    \includegraphics[width=\textwidth]{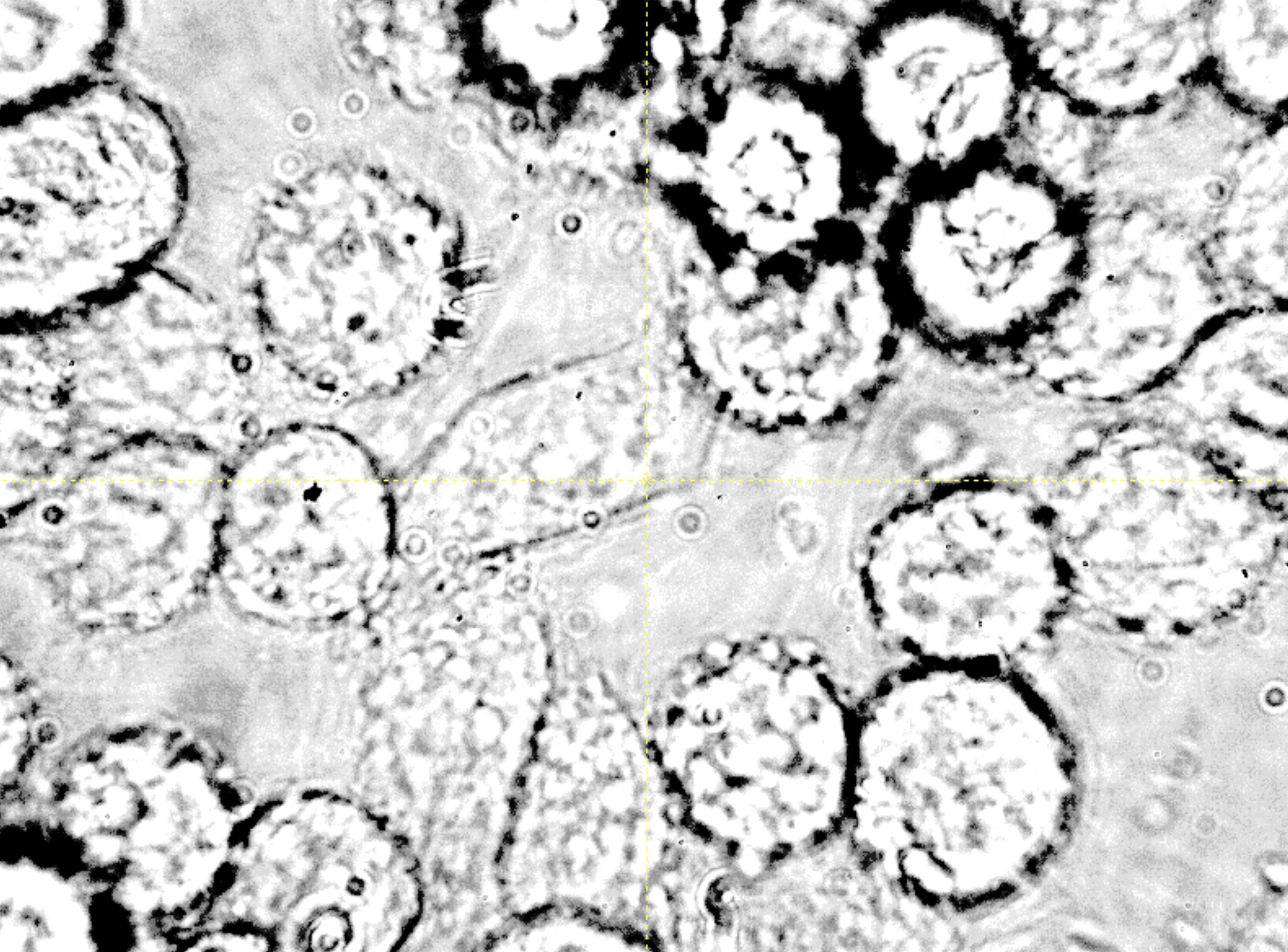}
    \caption{Brightfield image of the cell group}
    \label{fig:cells_pic}
  \end{subfigure}
  \hfill
\begin{subfigure}[t]{0.45\textwidth}
    \centering
    \includegraphics[width=\textwidth]{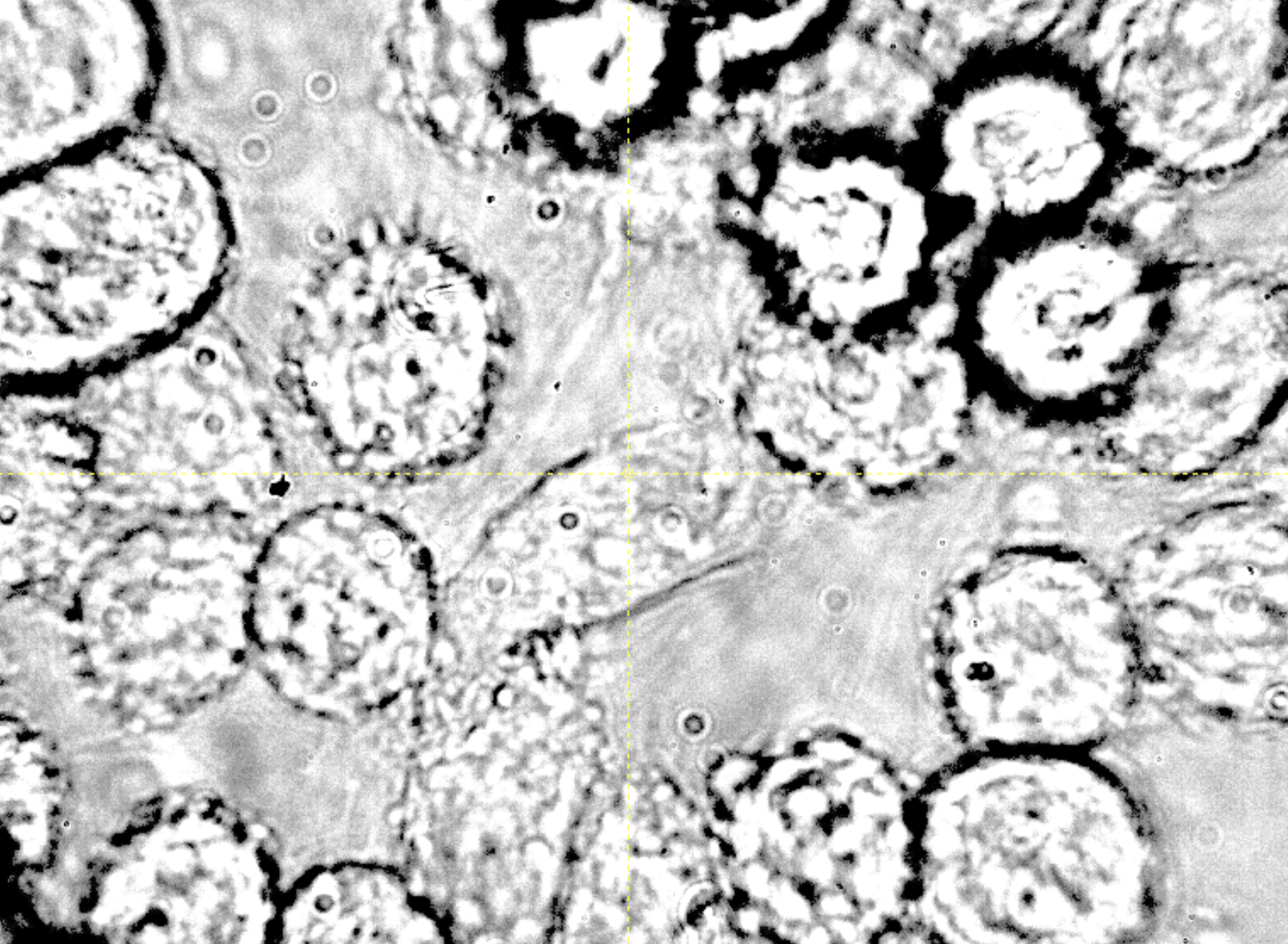}
    \caption{Brightfield image of the cell group after laser exposure}
    \label{fig:cells_pic_after}
\end{subfigure}
  \vspace{1em} 
  
  \begin{subfigure}[b]{0.45\textwidth}
    \centering
    \includegraphics[width=\textwidth]{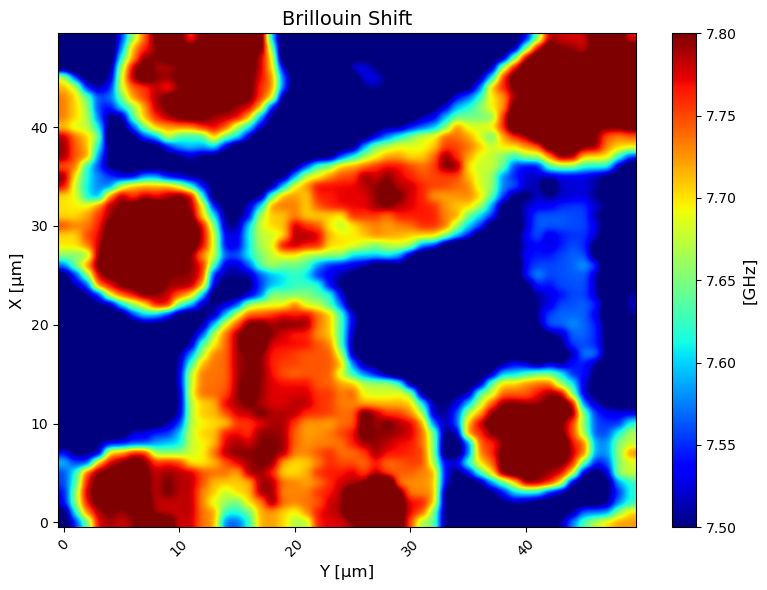}
    \caption{Brillouin shift heatmap}
    \label{fig:cells_shift}
  \end{subfigure}
  \hfill
  \begin{subfigure}[b]{0.45\textwidth}
    \centering
    \includegraphics[width=\textwidth]{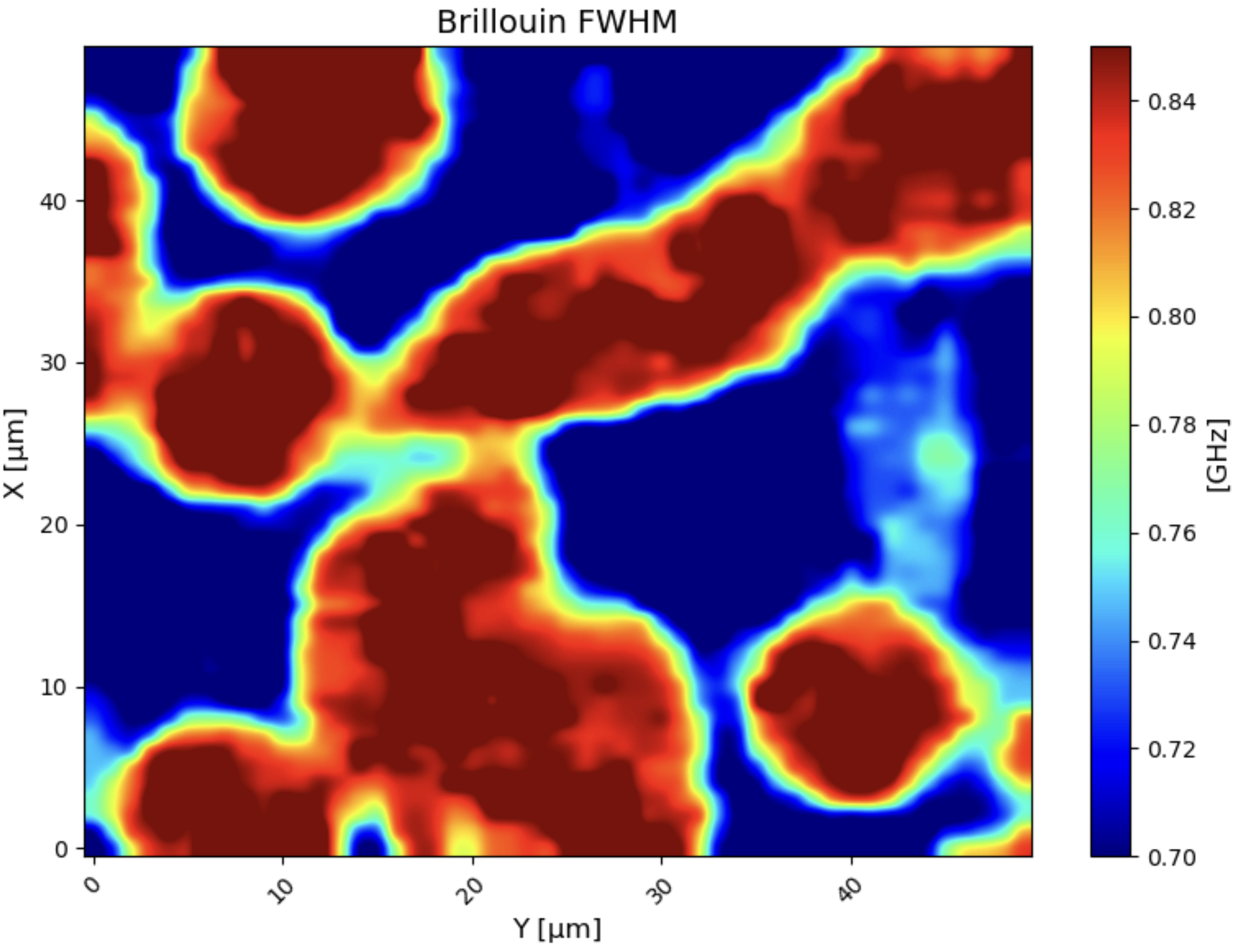}
    \caption{Brillouin FWHM map}
    \label{fig:cells_fwhm}
  \end{subfigure}
  
  \vspace{1em} 
  
  \begin{subfigure}[b]{0.45\textwidth}
    \centering
    \includegraphics[width=\textwidth]{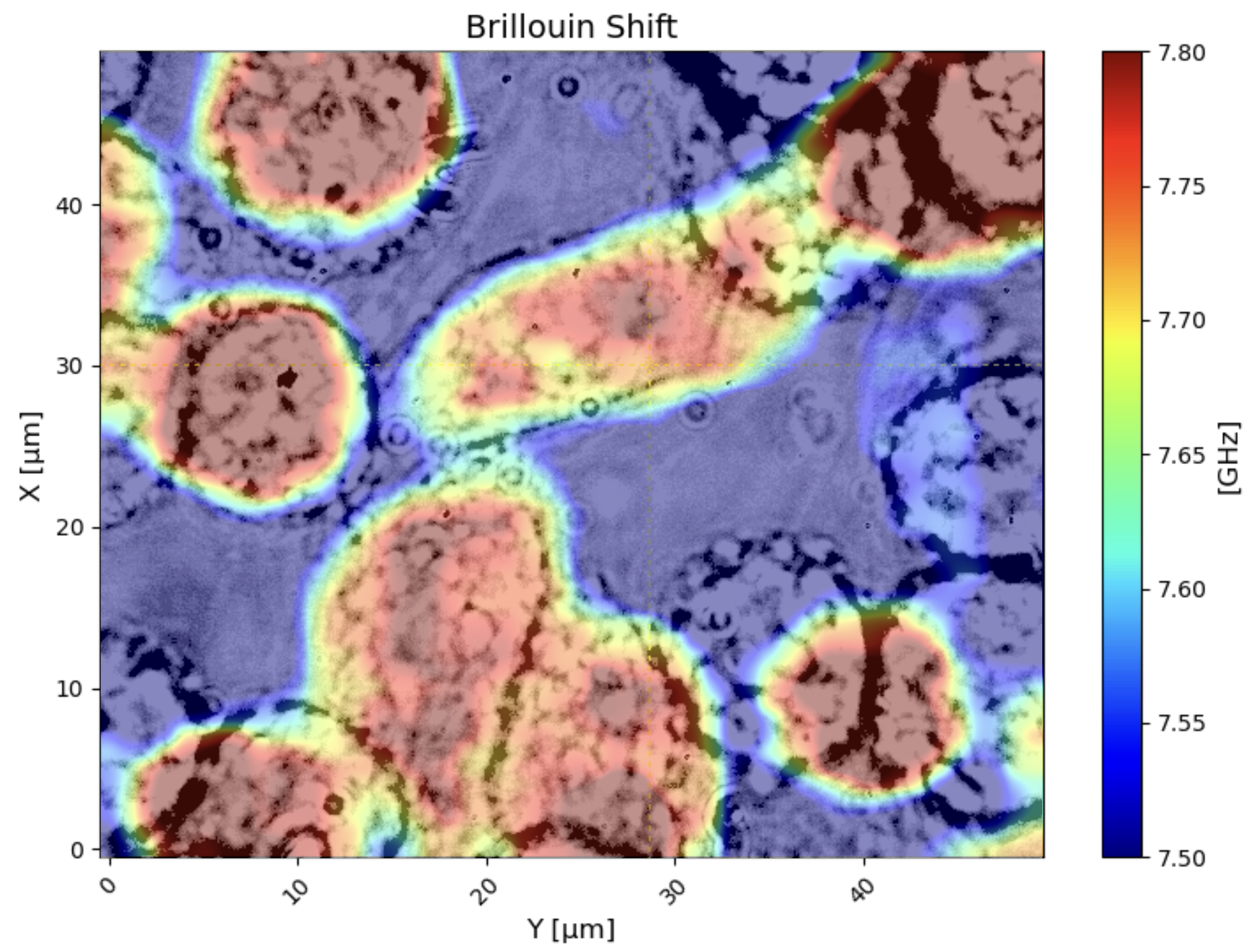}
    \caption{Overlay of Brillouin shift heatmap with cell image}
    \label{fig:cells_shift_overlay}
  \end{subfigure}
  \hfill
  \begin{subfigure}[b]{0.45\textwidth}
    \centering
    \includegraphics[width=\textwidth]{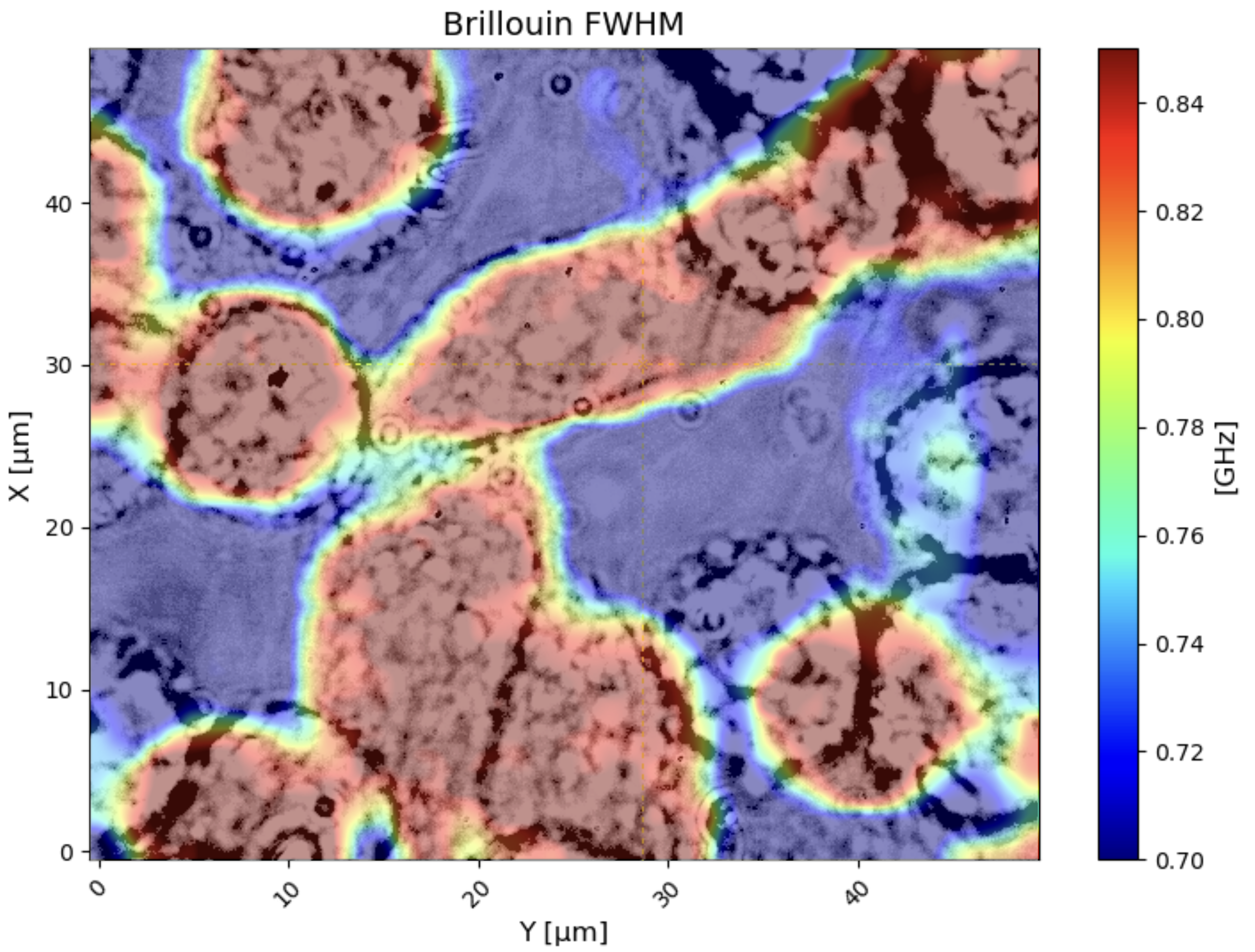}
    \caption{Overlay of Brillouin FWHM map with cell image}
    \label{fig:cells_fwhm_overlay}
  \end{subfigure}
  \caption{Processed Brillouin shift and FWHM heatmaps.}
  \label{fig:Brillouin_heatmap}
\end{figure}

For each pixel in each layer, the Brillouin spectrum was recorded. 
After recording each spectrum, we followed the next steps:
\begin{enumerate}
    \item The raw signal from the spectrometer's CCD was processed using a Savitzky-Golay filter \cite{savitzky1964smoothing} to enhance signal quality. The optimal signal-to-noise ratio (SNR) of the spectra was obtained with a filter window width of 7 and a fitted polynomial order of 3.
    \item The positions of specific spectral lines resulting from Rayleigh (elastically) scattered photons, as well as Stokes and Anti-Stokes Brillouin shifted lines, were precisely determined.

    \item Due to the non-linear spectral response of the Virtually Imaged Phase Array (VIPA), it was necessary to compute the dispersion curve of the spectrometer to accurately convert the observed shift values from the linear domain on the CCD sensor into the frequency domain. The dispersion curve was derived from the spectra corresponding to each measurement point, utilizing the fixed positions of spectral lines from Rayleigh (elastically) scattered photons. Given the operational principles of VIPA, we can determine the exact separation between these peaks in the frequency domain, which is established at 29.98 GHz.
    
    \item The values of Brillouin shift and Brillouin peaks Full Width Half Maximum (FWHM) are extracted from the spectra.
\end{enumerate}
The processed signal for a single pixel is illustrated in Fig. \ref{fig:Brillouin_spectrum}. The results of the processed signal from a single pixel are assembled into the heatmap as shown in Fig. \ref{fig:Brillouin_heatmap}.

\section{Results}
\begin{figure}[h!]
  \centering
  \begin{subfigure}[t]{0.45\textwidth}
    \centering
    \includegraphics[width=\textwidth]{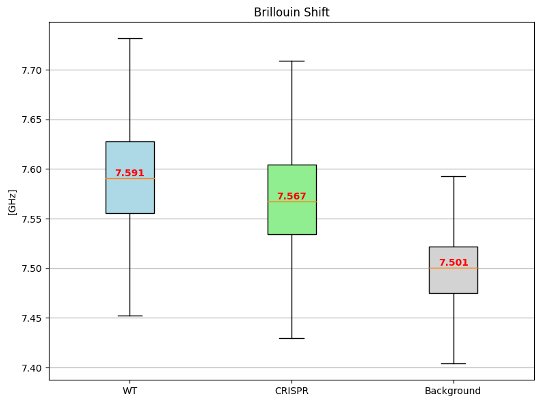}
    \caption{Brillouin shift box plot. WT median: 7.591 GHz; CRISPR median: 7.567 GHz.}
    \label{fig:shift_box}
  \end{subfigure}
  \hfill
  \begin{subfigure}[t]{0.45\textwidth}
    \centering
    \includegraphics[width=\textwidth]{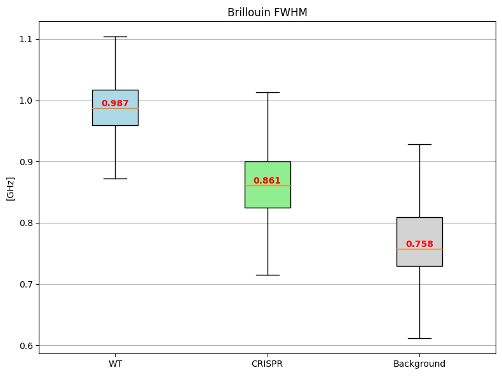}
    \caption{Brillouin FWHM box plot. WT median: 0.987 GHz; CRISPR median: 0.861 GHz.}
    \label{fig:fwhm_box}
  \end{subfigure}
  \caption{Box plots showing the distributions of (a) Brillouin shift and (b) FWHM values for wild-type (WT) and CRISPR-modified colon cancer cells.}
  \label{fig:box}
\end{figure}

Following the acquisition and pre-processing of the Brillouin spectra, we extracted the Brillouin shift and full width at half maximum (FWHM) values from each pixel. For quantitative analysis, twenty representative cells were arbitrarily selected from both the wild-type (WT) and CRISPR-modified groups. The regions corresponding to these cells were isolated, and the distributions of the Brillouin shift and FWHM values were compiled into box plots (see Fig. \ref{fig:box}).

As shown in Fig. \ref{fig:shift_box}, the box plot for the Brillouin shift reveals an median value of 7.591 GHz for the WT samples, while the CRISPR-modified cells exhibit a slightly lower median of 7.567 GHz. Similarly, Fig. \ref{fig:fwhm_box} shows that the median FWHM is 0.987 GHz for the WT cells, compared to 0.861 GHz for the CRISPR samples. These differences suggest subtle modifications in the viscoelastic properties between the two groups.

The observed decreases in both the Brillouin shift and FWHM in the CRISPR-modified cells may reflect alterations in the mechanical properties of the cellular structures, potentially linked to the genetic modifications. Further statistical analysis is underway to ascertain the significance of these differences and to correlate them with the underlying biophysical changes induced by CRISPR modification.

\section{Conclusion}
In summary, our study demonstrates that Brillouin spectroscopy can effectively resolve subtle differences in the viscoelastic properties of colon cancer cells. The quantitative analysis indicates that wild-type cells exhibit higher median Brillouin shifts (7.591 GHz) and broader FWHM values (0.987 GHz) compared to CRISPR-modified cells, which display medians of 7.567 GHz and 0.861 GHz, respectively. These results suggest that genetic modifications introduced via CRISPR may lead to measurable alterations in cellular stiffness and viscosity.

While these findings are promising, several factors warrant further investigation. The inherent heterogeneity within cells, potential stress effects from prolonged imaging, and the influence of averaging signals across the entire cell—particularly the contribution of the stiff nucleus—may all impact the measured parameters. Future work should aim to refine spatial resolution and isolate subcellular regions to better delineate these mechanical differences.

Overall, our results highlight the potential of Brillouin spectroscopy as a non-invasive diagnostic tool for differentiating cellular phenotypes in colon cancer. Continued optimization and validation of this technique could provide valuable insights into tumor mechanobiology and support the development of more targeted therapeutic strategies.

\bibliography{references}

\begin{thebibliography}{10}

\bibitem{araghi2019}
M.~Araghi, I.~Soerjomataram, A.~Bardot, J.~Ferlay, C.~Cabasag, D.~Morrison, P.~De, H.~Tervonen, P.~Walsh, O.~Bucher, et~al.
\newblock Changes in colorectal cancer incidence in seven high-income countries: a population-based study.
\newblock {\em The Lancet. Gastroenterology \& Hepatology}, 4(7):511--518, 2019.

\bibitem{oconnell2003}
J.~O’Connell, M.~Maggard, J.~H. Liu, D.~Etzioni, E.~Livingston, and C.~Ko.
\newblock Rates of colon and rectal cancers are increasing in young adults.
\newblock {\em The American Surgeon}, 69:866--872, 2003.

\bibitem{siegel2023}
R.~Siegel, N.~S. Wagle, A.~Cercek, Robert~A. Smith, and A.~Jemal.
\newblock Colorectal cancer statistics, 2023.
\newblock {\em CA: a cancer journal for clinicians}, 2023.

\bibitem{liss2014}
D.~Liss and D.~Baker.
\newblock Understanding current racial/ethnic disparities in colorectal cancer screening in the united states: the contribution of socioeconomic status and access to care.
\newblock {\em American Journal of Preventive Medicine}, 46(3):228--236, 2014.

\bibitem{singh2017}
Gopal~K. Singh and A.~Jemal.
\newblock Socioeconomic and racial/ethnic disparities in cancer mortality, incidence, and survival in the united states, 1950–2014: over six decades of changing patterns and widening inequalities.
\newblock {\em Journal of Environmental and Public Health}, 2017, 2017.

\bibitem{miller2019}
K.~Miller, Leticia~M. Nogueira, A.~Mariotto, J.~Rowland, K.~R. Yabroff, C.~Alfano, A.~Jemal, J.~Kramer, and R.~Siegel.
\newblock Cancer treatment and survivorship statistics, 2019.
\newblock {\em CA: A Cancer Journal for Clinicians}, 69, 2019.

\bibitem{egeberg2008}
R.~Egeberg, J.~Halkjaer, N.~Rottmann, Louise Hansen, and I.~Holten.
\newblock Social inequality and incidence of and survival from cancers of the colon and rectum in a population-based study in denmark, 1994-2003.
\newblock {\em European Journal of Cancer}, 44(14):1978--1988, 2008.

\bibitem{stock2010}
C.~Stock, U.~Haug, and H.~Brenner.
\newblock Population-based prevalence estimates of history of colonoscopy or sigmoidoscopy: review and analysis of recent trends.
\newblock {\em Gastrointestinal Endoscopy}, 71(2):366--381.e2, 2010.

\bibitem{mingsi2022}
Mingsi Wang, Yang Liu, Yi~Ma, Yue Li, Cheng-yao Sun, Yi~Cheng, Guoxiang Liu, and Xin Zhang.
\newblock Association between cancer prevalence and different socioeconomic strata in the us: The national health and nutrition examination survey, 1999–2018.
\newblock {\em Frontiers in Public Health}, 10, 2022.

\bibitem{seifeldin1999}
R.~Seifeldin and J.~J. Hantsch.
\newblock The economic burden associated with colon cancer in the united states.
\newblock {\em Clinical Therapeutics}, 21(8):1370--1379, 1999.

\bibitem{Alabi2019Decellularized}
Busola~R Alabi, R.~Laranger, and J.~Shay.
\newblock Decellularized mice colons as models to study the contribution of the extracellular matrix to cell behavior and colon cancer progression.
\newblock {\em Acta biomaterialia}, 2019.

\bibitem{Chen2016A}
H.~Chen, Zhubo Wei, Jian Sun, Asmita Bhattacharya, D.~Savage, R.~Serda, Y.~Mackeyev, S.~Curley, Pengcheng Bu, Lihua Wang, Shuibing Chen, L.~Cohen-Gould, E.~Huang, Xiling Shen, S.~Lipkin, N.~Copeland, N.~Jenkins, and M.~Shuler.
\newblock A recellularized human colon model identifies cancer driver genes.
\newblock {\em Nature biotechnology}, 34:845 -- 851, 2016.

\bibitem{Roper2017In}
J.~Roper, T.~Tammela, N.~Cetinbas, Adam Akkad, A.~Roghanian, S.~Rickelt, Mohammad Almeqdadi, Katherine Wu, M.~Oberli, Francisco~J. Sánchez-Rivera, Yoona~K Park, Xu~Liang, G.~Eng, Martin~S. Taylor, Roxana Azimi, Dmitriy Kedrin, R.~Neupane, S.~Beyaz, E.~Sicinska, Yvelisse Suarez, James Yoo, L.~Chen, L.~Zukerberg, P.~Katajisto, V.~Deshpande, A.~Bass, P.~Tsichlis, J.~Lees, R.~Langer, R.~Hynes, Jianzhu Chen, A.~Bhutkar, T.~Jacks, and Ömer H.~Yilmaz.
\newblock In vivo genome editing and organoid transplantation models of colorectal cancer.
\newblock {\em Nature biotechnology}, 35:569 -- 576, 2017.

\bibitem{Parsons2013Personalized}
B.~Parsons and Meagan~B Myers.
\newblock Personalized cancer treatment and the myth of kras wild-type colon tumors.
\newblock {\em Discovery medicine}, 15 83:259--67, 2013.

\bibitem{Yang1996Wild-type}
B.~Yang, J.~Eshleman, N.~Berger, and S.~Markowitz.
\newblock Wild-type p53 protein potentiates cytotoxicity of therapeutic agents in human colon cancer cells.
\newblock {\em Clinical cancer research : an official journal of the American Association for Cancer Research}, 2 10:1649--57, 1996.

\bibitem{Ogawa1997Novel}
N.~Ogawa, T.~Fujiwara, S.~Kagawa, M.~Nishizaki, Y.~Morimoto, T.~Tanida, A.~Hizuta, Tatsuji Yasuda, J.~Roth, and N.~Tanaka.
\newblock Novel combination therapy for human colon cancer with adenovirus‐mediated wild‐type p53 gene transfer and dna‐damaging chemotherapeutic agent.
\newblock {\em International Journal of Cancer}, 73, 1997.

\bibitem{Bouvet1998Adenovirus-mediated}
M.~Bouvet, L.~Ellis, M.~Nishizaki, T.~Fujiwara, Wenbiao Liu, C.~Bucana, B.~Fang, J.~Lee, and J.~Roth.
\newblock Adenovirus-mediated wild-type p53 gene transfer down-regulates vascular endothelial growth factor expression and inhibits angiogenesis in human colon cancer.
\newblock {\em Cancer research}, 58 11:2288--92, 1998.

\bibitem{Meng2023Application}
Hui Meng, Manman Nan, Yizhen Li, Y.~Ding, Yuhui Yin, and Mingzhi Zhang.
\newblock Application of crispr-cas9 gene editing technology in basic research, diagnosis and treatment of colon cancer.
\newblock {\em Frontiers in Endocrinology}, 14, 2023.

\bibitem{Sokhangouy2024Recent}
Saeideh~Khorshid Sokhangouy, Farzaneh Alizadeh, Malihe Lotfi, Samaneh Sharif, Atefeh Ashouri, Yasamin Yoosefi, Saeed~Bozorg Qomi, and M.~Abbaszadegan.
\newblock Recent advances in crispr-cas systems for colorectal cancer research and therapeutics.
\newblock {\em Expert review of molecular diagnostics}, pages 1--26, 2024.

\bibitem{Michels2020Pooled}
Birgitta~E. Michels, Mohammed~H. Mosa, Barbara~I. Streibl, T.~Zhan, C.~Menche, K.~Abou-El-Ardat, Tahmineh Darvishi, E.~Członka, Sebastian~A. Wagner, Jan Winter, H.~Medyouf, M.~Boutros, and H.~Farin.
\newblock Pooled in vitro and in vivo crispr-cas9 screening identifies tumor suppressors in human colon organoids.
\newblock {\em Cell stem cell}, 2020.

\bibitem{Hillman2023The}
Tatiana Hillman.
\newblock The use of plant-derived exosome-like nanoparticles as a delivery system of crispr/cas9-based therapeutics for editing long non-coding rnas in cancer colon cells.
\newblock {\em Frontiers in Oncology}, 13, 2023.

\bibitem{Drost2017Use}
J.~Drost, R.~van Boxtel, Francis Blokzijl, Tomohiro Mizutani, Nobuo Sasaki, Valentina Sasselli, J.~de~Ligt, S.~Behjati, Judith~E. Grolleman, T.~van Wezel, S.~Nik-Zainal, R.~Kuiper, E.~Cuppen, and H.~Clevers.
\newblock Use of crispr-modified human stem cell organoids to study the origin of mutational signatures in cancer.
\newblock {\em Science}, 358:234 -- 238, 2017.

\bibitem{Ramakrishna2021Application}
Gayatri Ramakrishna, Preedia~E. Babu, Ravinder Singh, and Nirupma Trehanpati.
\newblock Application of crispr-cas9 based gene editing to study the pathogenesis of colon and liver cancer using organoids.
\newblock {\em Hepatology International}, 2021.

\bibitem{Buckhaults2023Abstract}
P.~Buckhaults, Sanam Khalili, Carolyn~E. Banister, P.~Gokare, D.~Pocalyko, and K.~Bachman.
\newblock Abstract 251: Identification of therapeutic vulnerabilities by genome-wide crispr knockout library screening of colon cancer organoids.
\newblock {\em Cancer Research}, 2023.

\bibitem{Gao2021CRISPR}
Shanshan Gao, Fraser Soares, Shiyan Wang, Chi~Chun Wong, Huarong Chen, Zhenjie Yang, Weixin Liu, Minnie Y.~Y. Go, Musaddeque Ahmed, Yong Zeng, Catherine~Adell O’Brien, Joseph J.~Y. Sung, Housheng~Hansen He, and Jun Yu.
\newblock Crispr screens identify cholesterol biosynthesis as a therapeutic target on stemness and drug resistance of colon cancer.
\newblock {\em Oncogene}, 2021.

\bibitem{Sahranavard2023The}
T.~Sahranavard, S.~Mehrabadi, Ghazaleh Pourali, Mina Maftooh, H.~Akbarzade, S.~M. Hassanian, M.~Mobarhan, G.~Ferns, M.~Khazaei, and A.~Avan.
\newblock The potential therapeutic applications of crispr/cas9 in colorectal cancer.
\newblock {\em Current medicinal chemistry}, 2023.

\bibitem{Yin2019CRISPR}
Hao Yin, Wen Xue, and Daniel~G. Anderson.
\newblock Crispr–cas: a tool for cancer research and therapeutics.
\newblock {\em Nature Reviews Clinical Oncology}, 16:281--295, 2019.

\bibitem{Palombo2019Brillouin}
F.~Palombo and D.~Fioretto.
\newblock Brillouin light scattering: Applications in biomedical sciences.
\newblock {\em Chemical Reviews}, 119:7833 -- 7847, 2019.

\bibitem{Riobóo2021Brillouin}
R.~Riobóo, Nuria Gontán, Daniel Sanderson, M.~Desco, and M.~Gómez-Gaviro.
\newblock Brillouin spectroscopy: From biomedical research to new generation pathology diagnosis.
\newblock {\em International Journal of Molecular Sciences}, 22, 2021.

\bibitem{Troyanova-Wood2021Multi-Wavelength}
Maria~A. Troyanova-Wood and V.~Yakovlev.
\newblock Multi-wavelength excitation brillouin spectroscopy.
\newblock {\em IEEE Journal of Selected Topics in Quantum Electronics}, 27:1--5, 2021.

\bibitem{Ishijima2021Dispersive}
A.~Ishijima, Shinga Okabe, I.~Sakuma, and K.~Nakagawa.
\newblock Dispersive coherent brillouin scattering spectroscopy.
\newblock {\em Photoacoustics}, 29, 2021.

\bibitem{Ballmann2019Nonlinear}
C.~Ballmann, Zhaokai Meng, and V.~Yakovlev.
\newblock Nonlinear brillouin spectroscopy: what makes it a better tool for biological viscoelastic measurements.
\newblock {\em Biomedical optics express}, 10 4:1750--1759, 2019.

\bibitem{Seiler2019Brillouin}
T.~Seiler, Peng Shao, A.~Eltony, T.~Seiler, and S.~Yun.
\newblock Brillouin spectroscopy of normal and keratoconus corneas.
\newblock {\em American journal of ophthalmology}, 202:118--125, 2019.

\bibitem{2025CheburkanovBrillouinSPIE}
Vsevolod Cheburkanov, Mykyta Kizilov, Sujeong Jung, Mikhail~Y. Berezin, and Vladislav~V. Yakovlev.
\newblock Non-invasive remote assessment of tissue fibrogenesis using brillouin microscopy.
\newblock In Kirill~V. Larin and Giuliano Scarcelli, editors, {\em Optical Elastography and Tissue Biomechanics XII}, volume 13321, page 133210H. International Society for Optics and Photonics, SPIE, 2025.

\bibitem{2025CheburkanovGliaSPIE}
Vsevolod Cheburkanov, Mykyta Kizilov, Sujeong Jung, Karla I.~Ortega Sandoval, Shreya Raghavan, and Vladislav Yakovlev.
\newblock Noninvasive investigation of enteric glia culture viscoelastic properties.
\newblock In Natan~T. Shaked and Oliver Hayden, editors, {\em Label-free Biomedical Imaging and Sensing (LBIS) 2025}, volume 13331, page 1333106. International Society for Optics and Photonics, SPIE, January 26 2025.

\bibitem{Zhang2021Mapping}
Jitao Zhang and G.~Scarcelli.
\newblock Mapping mechanical properties of biological materials via an add-on brillouin module to confocal microscopes.
\newblock {\em Nature Protocols}, 16:1251 -- 1275, 2021.

\bibitem{Edrei2018Brillouin}
Eitan Edrei and G.~Scarcelli.
\newblock Brillouin micro-spectroscopy through aberrations via sensorless adaptive optics.
\newblock {\em Applied Physics Letters}, 112, 2018.

\bibitem{Li2021Quantum-enhanced}
Tian Li, Fu~Li, Xinghua Liu, V.~Yakovlev, and G.~S. Agarwal.
\newblock Quantum-enhanced stimulated brillouin scattering spectroscopy and imaging.
\newblock {\em Optica}, 9 8:959--964, 2021.

\bibitem{Elsayad2019Brillouin}
K.~Elsayad, F.~Palombo, T.~Dehoux, and D.~Fioretto.
\newblock Brillouin light scattering microspectroscopy for biomedical research and applications: introduction to feature issue.
\newblock {\em Biomedical optics express}, 10 5:2670--2673, 2019.

\bibitem{Troyanova-Wood2019Differentiating}
Maria~A. Troyanova-Wood, Zhaokai Meng, and V.~Yakovlev.
\newblock Differentiating melanoma and healthy tissues based on elasticity-specific brillouin microspectroscopy.
\newblock {\em Biomedical optics express}, 10 4:1774--1781, 2019.

\bibitem{Troyanova-Wood2016Elasticity-based}
Maria~A. Troyanova-Wood, Zhaokai Meng, and V.~Yakovlev.
\newblock Elasticity-based identification of tumor margins using brillouin spectroscopy.
\newblock 9719, 2016.

\bibitem{Cheburkanov2022Imaging}
Vsevolod Cheburkanov, Kavya Pendyala, Maria Parlani, T.~Lele, P.~Friedl, and V.~Yakovlev.
\newblock Imaging mechanical properties of cancer cells during metastasis with brillouin microspectroscopy.
\newblock 11944:119440C -- 119440C--9, 2022.

\bibitem{2025HarringtonDUVSPIE}
Joseph~T. Harrington, Vsevolod Cheburkanov, Mykyta Kizilov, Ilya Kulagin, Georgi Petrov, and Vladislav~V. Yakovlev.
\newblock Deep ultraviolet resonant raman (duvrr) spectroscopy for spectroscopic evaluation and disinfection of food and agricultural samples.
\newblock In {\em Photonic Technologies in Plant and Agricultural Science II}, volume 13357, page 1335702. International Society for Optics and Photonics, SPIE, 2025.

\bibitem{2025HarringtonDUVChemPhysChem}
Joseph Harrington, Vsevolod Cheburkanov, Mykyta Kizilov, Ilya Kulagin, Georgi~I. Petrov, and Vladislav~V. Yakovlev.
\newblock Highly sensitive, low-cost deep-uv resonant raman microspectroscopy systems.
\newblock {\em Chemistry–Methods}, n/a(n/a):2500006, 2025.

\bibitem{2025KizilovRamanSPIE}
Mykyta Kizilov, Vsevolod Cheburkanov, Joseph Harrington, and Vladislav Yakovlev.
\newblock Advanced preprocessing and analysis techniques for enhanced raman spectroscopy data interpretation.
\newblock In Robert~R. Alfano, Angela~B. Seddon, Lingyan Shi, and Binlin Wu, editors, {\em Optical Biopsy XXIII: Toward Real-Time Spectroscopic Imaging and Diagnosis}, volume 13311, page 133110F. International Society for Optics and Photonics, SPIE, 2025.

\bibitem{Rix2022Correlation}
J.~Rix, Ortrud Uckermann, K.~Kirsche, G.~Schackert, E.~Koch, M.~Kirsch, and R.~Galli.
\newblock Correlation of biomechanics and cancer cell phenotype by combined brillouin and raman spectroscopy of u87-mg glioblastoma cells.
\newblock {\em bioRxiv}, 2022.

\bibitem{Nikolić2018Noninvasive}
Miloš Nikolić, Christina Conrad, Jitao Zhang, and G.~Scarcelli.
\newblock Noninvasive imaging: Brillouin confocal microscopy.
\newblock {\em Advances in experimental medicine and biology}, 1092:351--364, 2018.

\bibitem{Moguilnaya2018Using}
T.~Moguilnaya, A.~Botikov, and A.~A. Agibalov.
\newblock Using coherent spectroscopy for diagnosing cancer at its early stages.
\newblock {\em Bulletin of the Russian Academy of Sciences: Physics}, 82:1052--1056, 2018.

\bibitem{savitzky1964smoothing}
Abraham Savitzky and Marcel~JE Golay.
\newblock Smoothing and differentiation of data by simplified least squares procedures.
\newblock {\em Analytical chemistry}, 36(8):1627--1639, 1964.

\end{thebibliography}

\end{document}